\documentclass[preprint,showpacs,showkeys,preprintnumbers,floatfix]{revtex4}
\usepackage{amsmath,epsfig}

\begin{document}

\preprint{SNUTP 03-005}

\title{Dynkin diagram strategy for orbifolding with Wilson lines}

\author{Kang-Sin Choi$^{(a,b)}$}
 \email{ugha@phya.snu.ac.kr}
\author{Kyuwan Hwang$^{(a)}$}
 \email{kwhwang@phya.snu.ac.kr}
\author{Jihn E. Kim$^{(a,b)}$}
 \email{jekim@phyp.snu.ac.kr}
\affiliation{$^{(a)}$School of Physics and Center for
Theoretical Physics, Seoul National University, Seoul
151-747, Korea,}
\affiliation{$^{(b)}$Physikalisches Institut,
Universit\"at Bonn, Nussallee 12, D53115, Bonn, Germany}

\begin{abstract}
A simple method for breaking gauge groups by orbifolding
is presented. We extend the method of Kac and Peterson to
include Wilson lines. The complete classification of the
gauge group breaking, e.g. from heterotic string,
is now possible. From this Dynkin diagram technique,
one can easily visualize the origin and the symmetry
pattern of the surviving gauge group.
\end{abstract}


\maketitle

\section{Introduction}

The heterotic string introduced an interesting
simply-laced semi-simple group $E_8\times E_8^\prime$ in
10 dimension(10D)
which can be a candidate for the fundamental theory
near the Planck scale~\cite{ghmr}. However, one must
compactify the six internal spaces so as to reconcile
with the observed 4 dimensional(4D) physics. Among a few
directions for compactification, the orbifolding has been
known to be especially simple and efficient, and
gives a rich spectra for 4D particles\cite{dhvw,inq}. In
addition, this orbifold method has led to interesting 4D
models which exhibit some desirable
physical phenomena such as the standard model gauge group
and the doublet-triplet splitting\cite{iknq}.

The so-called {\it standard-like} models, leading to the
gauge group $SU(3)\times SU(2)\times U(1)^n$ with
three chiral families, have been constructed vigorously
along this line\cite{iknq,munoz,imnq}. In case of no Wilson
line, there even exist extensive tables for all $Z_N$
orbifolds\cite{table}.
However, the possibilities for inclusion of Wilson
lines are exponentially larger than those without Wilson
lines. Therefore, the inclusion of Wilson line(s) was
limited to the study toward the standard-like models.
Along this line, an equivalence relation was even devised
to ease the study~\cite{cmmg}.

However, this initial study toward string
derivation of the supersymmetric standard model could
not overcome two serious hurdles along this
direction, one the $\sin^2\theta_W$ problem and the
other the problem of too many Higgs doublets.
Therefore, recently the orbifold has been tried at the
field theory level\cite{field}. Here, the doublet-triplet
splitting problem\cite{field} and the flavor
problem\cite{hwang} have been reconsidered, but it did not
exhibit a predictive power due to the arbitrariness of the
field content introduced at the fixed points.

This led us to consider the string orbifolds again. One
recent suggestion has been that a semi-simple gauge group
such as $SU(3)^3$ is plausible for a unification group
at high energy to solve the $\sin^2\theta_W$
problem in the orbifold compactification\cite{kim}.
This opens a new door toward string orbifolds.
In this respect, it is worthwhile to consider possible
string vacua through orbifold compactifications. Of course,
the previous classifications can be useful, but they are
not complete in that there does not exist a complete
classification with Wilson lines. Therefore, in this paper
we try to investigate how to implement the Wilson lines without
much computer time.

In this study, the simple group theoretical method, originally
devised by Kac and Peterson~\cite{kac}, is found to be extremely
useful. This method uses the Dynkin diagram, and can be understood
pictorially. In this method, we can easily tell the origin and the
symmetry of the resulting gauge group. However, cases with more
than one shift vector, which is the case of our interest with
Wilson lines, have not been studied completely. Here, generalizing
Kac and Peterson, we present a systematic search criteria for the
gauge groups in cases with more than one shift vector. And we seek
the behind symmetry of group, which is in fact apparent in the
Dynkin diagram. With the criteria
present in this paper, it is in principle possible to classify the
groups completely.

\section{Orbifold and shift vector}

\subsection{Breaking the group by a shift vector}

 We begin with the conventional root space the dimension of
which is the rank of the group.
Let us restrict the discussion to the {\em self
dual} lattices.
For a group $G$ and its root lattice spanned by
its roots $P$, we can make a transformation of $P$ by the shift
vector $V$,
\begin{align}
 |P\rangle & \mapsto e^{ 2 \pi i P \cdot V }|P\rangle, \label{charged} \\
 |Q\rangle & \mapsto |Q\rangle, \label{cartan}
\end{align}
where we have casted roots as states and $|Q\rangle$ is the set of
Cartan generators.
If we require that this transformation  is the symmetry of the system,
it breaks the group $G$ into its subgroup $H$, which consists of the root vectors
$|P\rangle$ satisfying
\begin{equation}
 P \cdot V = \mbox{integer}\,. \label{master}
\end{equation}
The {\em order} of the shift vector $V$ is defined
to be the minimum integer number $N$  such that $N$ successive
transformation becomes the identity operation up to a lattice
translation, i.e.,  $NV$ belongs to the root lattice.

\subsection{Orbifold}

The orbifold embedding of shift vector is a natural realization
of this property. An {\em orbifold} is defined by moding out the
manifold ${\bf R}^n$ by the space group ${\cal S}$,
which is seen equivalently as moding out the
torus $T^n$ by the {\em point group} ${\cal P}$,
$$
{\bf R}^n / {\cal S} = T^n / {\cal P}.
$$
Under ${\cal S}$, an element $x$ of ${\bf R}^n$ transforms as,
$$
x \mapsto \theta x + v.
$$
Here, the (usually rotation) element $\theta$ also
belongs to the point group ${\cal P}$, or the automorphism
of the lattice vector defining the torus. The order $N$
has the same meaning as that of the shift vector,
i. e., $\theta^N = 1$. Naively speaking, the
orbifolding is the identification of points on $T^n$ up to
the transformation $\theta$.  We associate this space group
transformation with the gauge group transformation,
\begin{align}
 \mbox{``rotation'' by } \theta & \to e^ {2 \pi i  P \cdot V}
\label{point} \\
 \mbox{translation by } v & \to e^{ 2 \pi i  P \cdot a}.
\label{transl}
\end{align}
We refer the latter as the Wilson line shift vector.

\subsection{Symmetries of Lie group}

A simple Lie group has many symmetries.
Any set of shift vectors which are connected by these
symmetries are equivalent, leading to the same subgroup.
\begin{enumerate}
\item {\it Lattice translation}: Although this is not a symmetry of
the gauge group, it is the redundancy of the formulation by the
  shift vector.
Under the translation
$$ V \mapsto V + \alpha, $$
by a root vector $\alpha$, the condition (\ref{master})
does not change, since the root vector is also an element of the dual lattice
(self dual),
any vector of which has only integer value when taking the dot product
with any root vector.

\item {\it Weyl reflection}: The Weyl
reflection($\sigma_\alpha$)
is defined as a reflection about the plane whose
normal vector is the root($\alpha$) of the group.
$$
V \mapsto \sigma_\alpha V = \sigma_{-\alpha} V
= V - 2\frac{\alpha\cdot V}{\alpha^2}\alpha
=V-(\alpha \cdot V)\alpha.
$$
We know that the set of any number of successive reflections form
a group. The group generated by the Weyl reflections is called
Weyl group.

\item {\it Outer automorphism}: The Dynkin diagram, which is a
diagrammatical representation of the Cartan matrix, contains
almost every information of the given group. Every small circle(or
bullet) represents a simple root and the linking lines
rerepresents the angle between the simple roots linked.[In our
case, we consider only small circles and singly connected lines.]
Some Dynkin diagram possesses some exchange symmetry, which is the
outer automorphism of the group. Some of the previous works
classifying orbifold models neglected this possibility. In fact,
many breaking patterns turn out to be the identical one.
\end{enumerate}

\section{Dynkin diagram technique for seaching
surviving groups}

What will be the subgroup $H$ which survives the condition
Eq.(\ref{master}) ?
The basic method of finding the group structure is to identify the
simple roots for the set of all roots which survive the projection
(\ref{master}).
By choosing the form of the shift vector carefully, one can find
the simple roots without identifying all roots surviving the projection.
For the case of one shift vector, i.e. when there is no wilson line
shift, Kac and Peterson~\cite{kac} introduced a very useful choice of
such form of the shift vector.

\begin{figure}[h]
\begin{center}
\epsfig{file=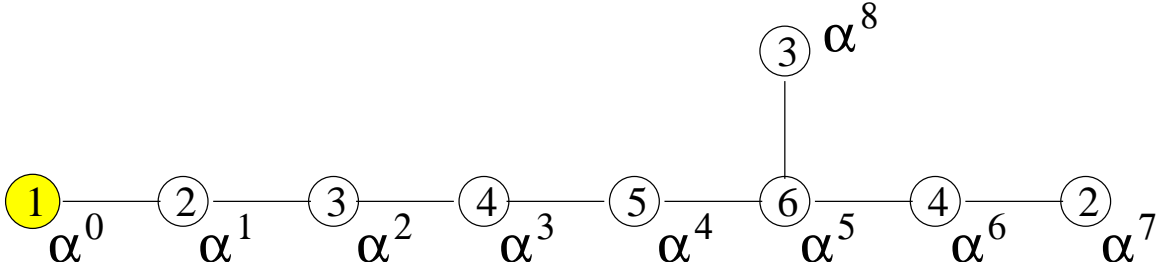}
\caption{  \label{E8dyn}
Extended Dynkin diagram of
$\widehat{E_8}$ group. The numbers in the circle are
the Coexter label $n_i$ of the corresponding simple roots.}
\end{center}
\end{figure}

 Let us concentrate on the group $E_8$. $E_8$ has eight simple
roots $\alpha^i, i=1..8$, represented by the small circles in the
(extended) Dynkin diagram in FIG.(\ref{E8dyn}).
The highest root $\theta$ of $E_8$ is given by
\begin{equation}\label{alpha0}
 \theta = 2\alpha^1 + 3\alpha^2
   + 4\alpha^3 + 5\alpha^4 + 6\alpha^5
   + 4\alpha^6 + 2\alpha^7 + 3\alpha^8.
\end{equation}
Let us define the {\it Coexter label} $\{n_i\}$ as
 the coefficients of $\alpha^i$ in this simple root expansion :
$$
\theta = \sum_{i=1}^r n_i \alpha^i,
$$
where $r={\rm rank~}{\rm of}~ G$.
If we define $\alpha^0 = -\theta$, its dot product with the other
simple roots vanishes except with $\alpha^1$, for which it is $-1$.
It cannot be included in the set of simple roots of $E_8$,
since $\alpha^0$ is not linearly independent on the other simple roots
$\alpha^i$ nor a positive root.
However, it can be a candidate of new simple root when some simple root
fails to pass the criterion Eq.(\ref{master}).
This set is called an extended root
system $\widehat{G}$ and the extended Dynkin diagram is shown in
FIG.(\ref{E8dyn}).

Let us expand the shift vector in the basis of the
fundamental weight, namely
{\em Dynkin basis}
$\{\gamma_i\}$, satisfying $\gamma_i \cdot \alpha^j
= \delta_i^j$,
\begin{equation}
 V = \frac 1 N \sum_{i=1}^r s_i \gamma_i, \label{shift}
\end{equation}
where $N$ is the order of the shift vector and $s_i$ is given by
$N \alpha^i \cdot V$.
We can limit our consideration for the shift vector satifying
\begin{equation}
\sum_{i=1}^{8} n_i s_i \le N, \quad s_i \ge 0\,, \label{lessn}
\end{equation}
for it is well known that one can always transform the shift
vector into this 'standard form' by lattice translation and Weyl
reflection of $E_8$.

With this form of shift vector, it is easy to identify the simple
roots of the surviving roots.
Let us define the set of indices $J$ which is made of the indices $i$
for which $s_i$ does not vanish.
\begin{equation}
J = \{i | s_i \ne 0\}
\end{equation}
First, let us consider the case where the inequality of Eq.(\ref{lessn})
holds.
For any positive root $P$ of $E_8$
\begin{equation}
P = \sum_{i=1}^{8} c_i \alpha^i\,, \quad c_i \ge 0\,,
\end{equation}
the projection condition Eq.(\ref{master}) becomes
\begin{equation}
P \cdot V = \sum_{i\in J} c_i s_i = {\rm integer}\,.
\end{equation}
This value is always smaller than one since that for the highest root
is less than one.
\begin{equation}
P \cdot V  \le \theta \cdot V = \sum_{i\in J} n_i \frac{s_i}{N} < 1
\end{equation}
Thus, any root having non-zero $c_i$ for $i \in J$,
fails to pass the condition. This means we can safely remove
the corresponding circle in the Dynkin diagram without causing any
side effect. The resulting subgroup can be read off from the resulting diagram.

Now, let us consider the case when the equality in Eq.(\ref{lessn}) holds.
In this case, there are some roots,
which still survives the projection condition even though $c_i \ne 0$ for
some $i\in J$,
which means we cannot just remove
the corresponding circle in the Dynkin diagram.
Let us call the set of such roots as $A$.
Any positive root $\tilde{P}$ in $A$ has $c_i = n_i$ for all indices $i \in J$,
hence it is expressed as
\begin{equation}
\tilde{P} = \theta - \sum_{i \not\in J} b_i \alpha^i
= -\alpha^0 - \sum_{i \not\in J} b_i \alpha^i\,,
\quad b_i \ge 0\,,
\end{equation}
for some non-negative integer $b_i$.
Hence any negative root $-\tilde{P}$, being the minus of $\tilde{P}$, in $A$ can be
expressed by the sum of the root vectors $\{\alpha^0$, $\alpha^i$
for $i\not\in J\}$ with positive coefficients.
Thus, by changing the definition of positive root for the root in $A$
in such a way that a previous negative root is a positive root and
vise versa, one can confirm that \{$\alpha^0$,  $\alpha^i$ for
$i \not\in J$\} forms a proper set of simple roots.
Thus the surviving subgroup is represented by the extended Dynkin
diagram with $i$-th circle being removed for $i \in J$, i.e. $s_i \ne 0$.

Observing that whether we add the extended simple root $\alpha^0$ to the
Dynkin diagram or not depends on the value $s_0$ defined by
\begin{equation}
s_0 \equiv N(1 - \theta\cdot V) = N - \sum_{i=1}^{8} n_i s_i
\end{equation}
vanishes or not, both cases considered in the last two paragraphs
can be treated universally by formaly defining the Coexter label $n_0$
for the extended simple root $\alpha^0$ as 1 and restrict our
consideration to the case which satisfies
\begin{equation}  \label{Kacond}
\sum_{I = 0}^{8} n_I s_I = N\,, \quad s_I \ge 0\,.
\end{equation}
For a solution $\{s_I, I=0..8\}$ of Eq.(\ref{Kacond}),
the unbroken subgroup by the shift vector given by Eq.(\ref{shift})
can be read off
from the extended Dynkin diagram with $I$-th circle removed for
any index $I$ with $s_I \ne 0$ including $I=0$.
If the rank of the diagram is smaller than 8, there are additional
$U(1)$'s which fill up the rank to 8.
The trivial case is $ s_0 = N$ where the
group $G$ remains unbroken.

\bigskip

\begin{figure}[h]
\begin{center}
\epsfig{file=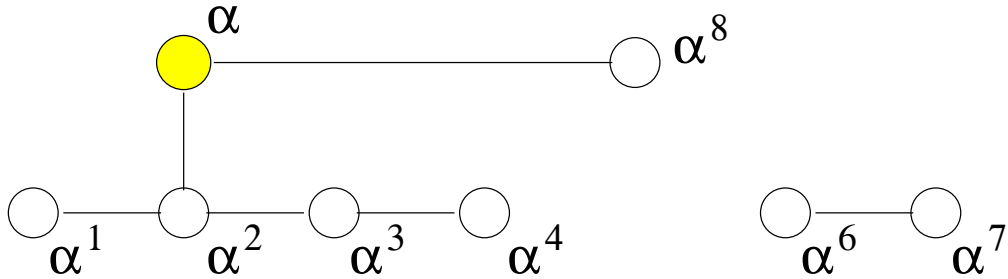}
\caption{ \label{odde6}
An order $N=3$ breaking. A
shift vector which deletes the 5$^{\rm th}$ spot
leads to $E_6 \times
SU(3)$. Compare this with FIG. 1.}
\end{center}
\end{figure}

It will be interesting to think about what will be the unbroken subgroup
for the shift vector which does not satisfy the condition Eq.(\ref{lessn}).
For example, take $V=\frac 13\gamma_5$,
hence $\sum n_i s_i= 6 > N $ where $N=3$.
It is tempting to speculate that deleting the 5th
circle($\alpha_5$) in the extended Dynkin diagram
resulting in $SU(6) \times SU(2) \times SU(3)$.
However, the set $\{\alpha^0, \alpha^i\} - \{\alpha^5\}$ cannot
generate a root vector with $c_5 = 3$ which survives the projection.
In this case, the proper simple root turns out to be
\def\hf{\frac12}
$$
\alpha
= \alpha^3 + 2\alpha^4 + 3 \alpha^5 + 2\alpha^6
+ \alpha^7 + \alpha^8\,.
$$
It links to $\alpha^2$ and $\alpha^8$,
whose resulting ``extended'' Dynkin
diagram is again $E_6\times SU(3)$, which
is depicted in FIG. \ref{odde6}.
By the theorem above, there is an equivalent shift vector
obeying the condition (\ref{lessn}) leading to the same subgroup.
One can show that this is an equivalent breaking to
the second one in TABLE \ref{Z3tbl}, by lattice translations and Weyl
reflections.

The complication of this example comes from the fact that there are
surviving roots which cannot be expressed in terms of the highest root
and the surviving simple roots. Those roots are allowed since
$\theta \cdot V = \sum n_i s_i/N$ is greater than 1.
Hence this example
shows the importance of the condition Eq.(\ref{lessn}) in reading off
the gauge group directly from the Dynkin diagram.

\bigskip

The power of the condition Eq.(\ref{Kacond}) lies not only in the
easyness to identify the unbroken subgroup but also in the fact that
there are only a few possible set of $s_I$'s that can satisfy it.
Thus the procedure of finding all possible gauge symmetry breaking
by the shifts can be very simplified using the condition Eq.(\ref{Kacond})
and the extended Dynkin diagram.
For example, let us consider an order $N=3$ shift.
From Eq.~(\ref{Kacond}) and watching
$n_I$, there are only five possibilities.
The shift vector~(\ref{shift}) in the Dynkin basis and the corresponding
five unbroken groups are listed in TABLE \ref{Z3tbl}.

\begin{table}
\begin{center}
\begin{tabular}{cc}
\hline
$[s_i|s_0]$ & unbroken group $H$ \\
\hline
$[00000000|3]$ & $E_8$ \\
$[01000000|0]$ & $E_6 \times SU(3)$ \\
$[00000001|0]$ & $SU(9)$ \\
$[10000000|1]$ & $E_7 \times U(1)$ \\
$[00000010|1]$ & $SO(14) \times U(1)$ \\
\hline
\end{tabular}
\caption{
\label{Z3tbl}
The order $N=3$ breakings of $E_8$.
The last entry $s_0$ can
be added automatically to satisfy Eq.~(\ref{Kacond}).}
\end{center}
\end{table}

\section{Introduction of Wilson line}

In addition to the shift vector associated with the point group,
we introduce another shift vector when we turn on a Wilson
line~\cite{inq}  $a$, which should also satisfy the same condition
as $V$ and is required to satisfy additional conditions on the
modular invariance stated in the next section. Extending the
diagramatic method of finding the unbroken subgroup for the first
breaking of $E_8$, we can find a well-defined procedure of finding
the unbroken subgroup for the further breaking by the additional
Wilson line shift vector. One can introduce as many Wilson lines
as the number of the compact dimensions in the orbifold
compactification bases on torus. Most of the statements in this
section assumes that the procedure is applied recursively for each
additional Wilson line $a_i$. Some of the statements is for the
first Wilson line $a_1$, for the sake of definiteness and
simpleness of the argument, though. These statements can be
generalized appropriately to the next Wilson line easily.

 Let us call the unbroken subgroup of the first breaking in the
previous section as $H = H_1 \times H_2 \times \dots$, where each $H_x$
is a simple group.
  The first step of the method in the previous section was to find an
efficient form of a shift vector expanded in the dual basis, i.e.
the fundamental weight of $E_8$.  For now we are dealing with the
subgroup of $E_8$, one might try to expand in terms of the
fundamental weights of $H_x$.  However, it is not useful since
 the fundamental weight vectors
 of $H_x$ are not on the root lattice of $H_x$
since it does not form a self-dual lattice in general.
For some cases, they reside in the root lattice of $H_x$ if they are
multiplied by some positive integer $\hat{N}$, implying they might be useful
for $Z_{\hat{N}}$ orbifold. Still, it lacks the generalities and do not
allow the formulation for the general order $N$, hence we do
not use them in this paper.
Instead, we keep using the expansion by the fundamental weights of $E_8$,
$\gamma_i$.
They are not exactly a dual of the simple roots we found in the
previous section in the strict sence, since $\alpha^0$, the negative
of the highest root of $E_8$, which does not have its dual among $\gamma_i$,
can be one of the simple roots of $H$.
However, this expansion is quite useful since all the other
simple roots of $H$ are still identified as simple roots
of $E_8$.
For the simple root $\alpha^0$, if present, whether the corresponding
circle should be removed or not can be deduced from the coefficients of
the other simple roots.
 The central point of our method is to provide a neat formulation of
keeping track of the fate of those extended simple roots at the
previous stage in a similar form of Eq.(\ref{Kacond}).

We start our discussion by
defining the highest root $\theta^x=-\alpha^{H_x}$ and
the labels $\{n^{H_x}_I\}$ for each simple group factor $H_x$ as
\begin{equation}
 \theta^x = \sum_{i \in J_x} n_i^{H_x} \alpha^i\,,
\end{equation}
where $J_x$ is the set of indices $i$ for which
the $E_8$ simple root $\alpha^i$, including $\alpha^0$,
is identified to be a simple root of
 $H_x$ at the previous stage of breaking.
By adding $\alpha^{H_x}$, the root system of each $H_x$
is extended to $\widehat{H_x}$.
We expand the Wilson line shift vector $a$ in terms of
the fundamental weights of $E_8$.
\begin{equation}
 a = \frac 1N \sum_{i=1}^{8} w_i \gamma_i\,.
 \label{wilshift}
\end{equation}
 The next important step is to reduce the possibilities of $w_i$ by
requiring the analogous condition of Eq.(\ref{lessn}). For each
simple group $H_x$, we can transform a shift vector $a$ into its
own standard form by the lattice translation and the Weyl
reflection by the root of $H_x$. Here we can not use the standard
form itself since we do not expand $a$ in terms of the fundamental
weight of $H$. Still, a slightly different form of the same
theorem is very useful to constrain the shift vector $a$:

\medskip

{\it Theorem :} One can always transform any vector $a$ in the
combined root space of $H=H_1\times H_2\times ...$ by lattice
translation and Weyl reflection by the root of $H_x$ into the one
that satisfies
\begin{equation}
N \theta^x \cdot a = \sum_{i\in J_x} n_i^{H_x}
\left(N \alpha^i \cdot a\right) \le N,
\quad\quad \alpha^i \cdot a \ge 0
\end{equation}
for each simple group $H_x$. This transformation does not change
the set of shift vectors $\{V,a_1, ...\}$ which leads to the
symmetry breaking $E_8 \to H$, upto lattice translation.

\medskip

The proof of the first statement of the theorem is manifest
since it is just the
substitution of $s_i$ by $N\alpha^i \cdot a$ in Eq.(\ref{lessn})
and each simple group $H_x$ is completely independent among one another
since $\alpha^i \cdot \alpha^j = 0$ if $\alpha^i$ and $\alpha^j$ belong
to different subgroup.
 The second statement of the theorem, which is an important consistency
condition of our scheme, is also obvious since any root $P$ which
survive the projection of the previous stage satisfies $P \cdot V
=$ integer, $P \cdot a_1=$ integer, etc..

For the sake of simplicity, we illustrate the usage of this theorem in
dealing with the extended root $\alpha^0$ of $E_8$ only.
The generalization to the other extended roots for the subgroup of
$E_8$ can be made easily.
 For the simple group $H_x$ which does not contain $\alpha^0$ as its
simple root, this theorem is simply equivalent to
\begin{equation} \label{lessn_sub_n}
\sum_{i \in J_x} n_i^{H_x} w_i \le N\,, \quad w_i\ge 0\,,
\end{equation}
just like the $E_8$ case in the first breaking by $V$.
 For the simple group $H_x$ which contains $\alpha^0$,
it translates into
\begin{eqnarray}  \label{lessn_sub_s}
N\theta^{x} \cdot a &=& \sum_{i\in J_x - \{0\}} n_i^{H_x} w_i
  \ \ +\  n_0^{H_x} w_0\ \ \le \ N \,,  \quad\quad
w_i, w_0 \ge 0\,,  \\
w_0 &\equiv& N\alpha^0 \cdot a = -\sum_{i=1}^{8} n_i w_i\,.
\label{w0}
\end{eqnarray}
The non-trivial condition on the quantity $w_0 \ge 0$ is the simple consequance
of the theorem. The apparent sign inconsistency between Eq.(\ref{w0}) and
Eq.(\ref{lessn_sub_n}),(\ref{lessn_sub_s}) is to be resolved by the
negative value of $w_i$ with $i \in J$, where $J$ is the set of all indices
$i$ for which $\alpha^i$ fails to pass the projection at the previous stage.
Thus  $w_i$ for $i\in J$ should be fixed by hand in such a way that
$w_0$ defined by Eq.(\ref{w0}) satisfies Eq.(\ref{lessn_sub_s}).
When $J$ has only one element $\hat{i}$, the constraints on $w_0$ is
equivalent to find a solution of
\begin{equation}  \label{Kacond_w0}
\sum_{I = 0}^{8} n_i w_i = N \quad \mbox{mod}\ N\,,
\end{equation}
with $w_{\hat{i}} = 0$ since $n_{\hat{i}}$ is already a multiple
of $N$. This property is useful for some physically interesting
cases like further breaking of $SU(3)\times E_6$ on $Z_3$ orbifold
which we will see in detail in the next section. When $J$ has more
than one element, there may be more than one way for $w_i, i\in J$
to satisfy Eq.(\ref{w0}). Each such solution may or may not lead
to the same gauge group, depending on whether they are related to
one another by Weyl reflection or not. It cannot be knwon at this
stage in general way. Thus, for the classification purpose, they
must be treated as a different one unless they are proven to be
related by Weyl reflection.

Given the restriction of Eq.(\ref{lessn_sub_n}),(\ref{lessn_sub_s}),
with the proper definition of $w_0$ in Eq.(\ref{w0}), we can define
the coefficient $w_0^{H_x}$ for the extended root $\alpha^{H_x}$
of the simple group factor $H_x$ such that
\begin{equation} \label{Kacond_sub}
\sum_{I\in J_x} n_i^{H_x} w_i + w_0^{H_x} = N\,, \quad
w_i, w_0^{H_x} \ge 0\,,
\end{equation}
by the same argument of the previous section.
One can read off the surviving gauge group from the extended Dynkin
diagram after removing of the corresponding circle for
$\alpha_i$ or $\alpha^{H_x}$ if
$w_i\ne 0$ or $w_0^{H_x} \ne 0$.

\medskip

If there are more than one Wilson line, we can do this procedure
recursively with every extended root $\alpha^{H_x}$
at the previous stage being taken care of by the same way as $\alpha^0$
is in the above illustration.
As we took more Wilson lines, surviving gauge group has
more factor groups with smaller rank, leading to many extended
roots to be taken care of. It may become quite tedious but is
trivial at the same time since we deal with small groups like
$SU(n)$ for which all Coexter label is one.

\subsection{Further breaking of $SU(3)\times E_6$}

\begin{figure}[h]
\epsfig{file=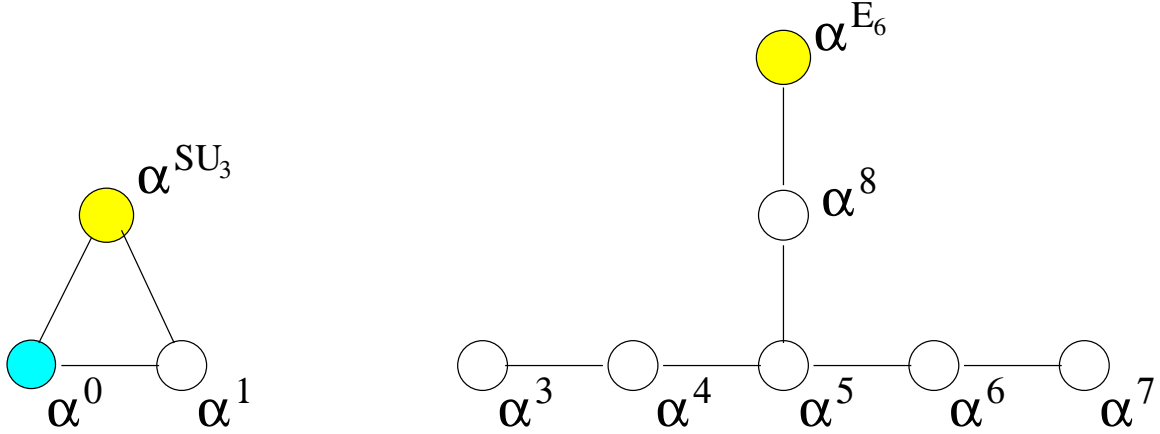}
\caption{
\label{e6dyn}
Extended Dynkin diagrams for
$\widehat{SU(3)}$ and $\widehat{E_6}$. New labels are defined for
each subgroup. Compare these with FIG. 1.}
\end{figure}

As an illustration of the method presented in the previous section.
we will study the further breaking of $SU(3)\times E_6$ by one
Wilson line.  The subgroup $SU(3)\times E_6$ is obtained when the
orbifold shift vector is given by $V= \gamma_2/3$ for $Z_3$ orbifold,
as can be seen from the TABLE \ref{Z3tbl}.
The $E_6$ has $\alpha^3, \cdots, \alpha^8$ as its simple roots
and the highest weight is given by
\begin{equation}
\theta^{E_6} = -\alpha^{E_6} = \alpha^3 +
2\alpha^4 + 3 \alpha^5 + 2\alpha^6 + \alpha^7 + 2 \alpha^8\,.
\end{equation}
On the $SU(3)$ side, the simple roots consists of $\alpha^1$ and
$\alpha^0$, the
$E_8$ extended root,  and the highest weight is given by
\begin{equation}
\theta^{SU_3} = -\alpha^{SU_3} = \alpha^0 + \alpha^1\,.
\end{equation}
The extended Dynkin diagrams for $SU(3)\times E_6$ are shown in
the FIG. \ref{e6dyn}.

\begin{enumerate}
\item
Take $a=\frac13 \gamma_4$, or
$w=[00010000|w_{0} w_0^{SU_3} w_0^{E_6}]$.
Since $\sum n_i w_i = 5 \ne 0$ mod 3, $\alpha^0$ fails to pass the projection.
$w_0$ is determined to be 1 with $w_2 = -2$ from Eq.(\ref{w0}).
Having
$n_0^{SU_3}=1$ and $n_4^{E_6} = 2$,
the other coefficients are determined to be
$w_0^{SU_3} = 2$ and $w_0^{E_6} = 1$
from Eq.(\ref{lessn_sub_n}) and (\ref{lessn_sub_s}).
The resulting group is
$SU(2)\times SU(2)\times SU(5)$ and the final
$w$ leading to this group is $w=[00010000|121]$.
We leave $w_2$ to be zero for simplicity in this paper since it gives
no information about the group structure.

\item
Consider $a=\frac 13 \gamma_8 $. In this case $\alpha^0$ survives
since $\sum n_iw_i=3$. Thus no simple root in $SU(3)$ group fail to
pass the projection. We can take either $w_0 =0$ or $w_0 =3$, the
result is the same, being $w_0^{SU_3}=3$ or $w_0^{SU_3}=0$, respectively.
For $E_6$ part, $n_8^{E_6}=2$, thus we have $w_0^{E_6}=1$.
As a result, we have $w=[00000001|031]$ and the resulting group
is $SU(3)\times SU(6)$.

\item
A more interesting case is $a=\frac13 \gamma_5$, or $w=[00001000|030]$.
Both $\alpha^0$ and $\alpha^{E_6}$ survive.
Therefore, a fairly large group of $SU(3)^4$ survives.
By looking at the Dynkin diagram, one notes
a symmetry by permutating three of $SU(3)$'s.
This has been used for an $SU(3)^3$ unification
model\cite{kim}.

\end{enumerate}

In this way, all possible other $w_i$'s are examined and the resulting gauge
groups are listed in TABLE \ref{e6furth}.
Some comments are in order.
 The extended Dynkin diagram of $\widehat{E_6}$
shows the symmetry, namely {\em triality}, under permutations of
$(\alpha^3,\alpha^4), (\alpha^7,\alpha^6)$ and
$(\alpha^{E_6},\alpha^8)$ pairs. Nonetheless, the breakings of
$E_6$ by deleting the corresponding spots under this triality are
not Weyl-equivalent since they have different $E_8$ characteristic
like the Coexter labels $\{n_i\}$.
 As a result, it can be easily seen that some $w$ vectors
resulting in the same gauge group are
related by the triality symmetry.

\begin{table}
\begin{center}
\begin{tabular}{cc}
\hline
$[w_i|w_0 w_0^{SU_3} w_0^{E_6}]$ & group \\
\hline
$[10000000|113]$ & $E_6$ \\
$[00100000|212]$ & $SU(2) \times SO(10)$ \\
$[00000010|122]$ & $SU(2) \times SO(10)$ \\
$[00010000|121]$ & $SU(2)^2 \times SU(5)$ \\
$[00000100|211]$ & $SU(2)^2 \times SU(5)$ \\
$[00000001|031]$ & $SU(3)\times SU(6)$ \\
$[00001000|000]$ & $SU(3)^4$ \\
\hline
$[10100000|022]$ & $SU(2)\times SO(10)$ \\
$[10000010|202]$ & $SU(2)\times SO(10)$ \\
$[10010000|201]$ & $SU(2)^2\times SU(5)$ \\
$[10000100|021]$ & $SU(2)^2\times SU(5)$ \\
$[10000001|111]$ & $SU(6)$ \\
$[10001000|110]$ & $SU(3)^3$ \\
$[00110000|030]$ & $SU(3)\times SU(6)$ \\
$[00000110|030]$ & $SU(3)\times SU(6)$ \\
$[00100100|120]$ & $SU(2)^2\times SU(5)$ \\
$[00010010|210]$ & $SU(2)^2\times SU(5)$ \\
$[00100001|210]$ & $SU(2)^2\times SU(5)$ \\
$[00000011|120]$ & $SU(2)^2\times SU(5)$ \\
$[00100010|031]$ & $SU(3)\times SO(8)$ \\
\hline
$[10110000|110]$ & $SU(6)$ \\
$[10000110|110]$ & $SU(6)$ \\
$[10100100|200]$ & $SU(2)^2\times SU(5)$ \\
$[10010010|020]$ & $SU(2)^2\times SU(5)$ \\
$[10100001|020]$ & $SU(2)^2\times SU(5)$ \\
$[10000011|200]$ & $SU(2)^2\times SU(5)$ \\
$[10100010|111]$ & $SO(8)$ \\
\hline
\end{tabular}
\end{center}
\caption{ \label{e6furth} Possible further breakings of $E_6
\times SU(3)$. $U(1)$ factors are implied to make the rank 8. Note
that $\sum n_I w_I = 0$ mod 3, $\sum n_i^{E_6} w_i = 3, \sum
n_i^{SU_3} w_i = 0$. The shift vectors resulting in the same group
are related by a symmetry. We do not list the
cases with some $w_i$'s of 2, because they do not
provide any new symmetry breaking patterns.
 }
\end{table}

\section{Heterotic string and modular invariance}

So far the procedure presented for finding an unbroken subgroup
of $E_8$ by the Dynkin diagram has been general.
In string theory, however, the consistency under the quantum
corrections demands further conditions which is called the
modular invariance conditions. In this case, the
unbroken group by the shift vectors is more restricted.
The search for the groups via shift vectors was the original
motivation for introducing the orbifold
compactification~\cite{dhvw} in physics. In fact, these
root systems naturally arise from the heterotic string~\cite{ghmr}.
Here, we employ the bosonic string description. A string state
excited by an oscillator describes a Cartan generator in
Eq.~(\ref{cartan}). The momentum and winding states around
the compact
dimension, which describe ``charged bosons'' under this
Cartan generator
describe roots in Eq.~(\ref{charged}). The modular invariance
condition restricts the relations among these states.

By orbifolding we identify the space (in the compact dimension
coordinate $z_m=x_{2m}+i x_{2m+1}$)
$$ z_m \sim z_m e^{ 2 \pi i \phi_m }  $$
Similarly shift vector is defined on the group space
$$ z_n \sim z_n e^{ 2 \pi i v_n } $$
By the definition of order, $N$ successive twist is identity
operation, hence
$$
\sum N \phi_m = \sum N v_n = 0 \quad (\rm{mod}~2),
$$
where the modulo 2 condition results for the case of
spinorial states, e.g. in the $E_8\times E_8^\prime$ theory.

Here, the modular invariance condition~\cite{dhvw} for
the orbifold restricts the form of the shift vector.
Constraining the discussion on
the abelian orbifold, i.e. when the orbifold action is
commutative~\cite{imnq} then the only necessity from the modular
invariance condition~\cite{dhvw} is that under $\tau \mapsto \tau
+1$. In terms of the shift vectors $\phi$ and $V$, we have
\begin{align}
(N\phi)^2 &= (N V)^2, \quad (\mbox{mod } 2N)\label{modinv}
\end{align}
where
\begin{align}
(N V)^2 &= ( \sum s_i \gamma_i )^2 \notag \\
          &= \sum_{ij} s_i  A^{-1}_{ij} s_j \label{modinv1} \\
          &= s \cdot A^{-1} s, \notag
\end{align}
where the dot product between two vectors in the Dynkin basis is
understood. In this basis, we can easily read off the modular
invariance condition from the inverse Cartan matrix $A^{-1}$.
This is especially useful in
implementing the procedure into the computer code.
For a concrete form of the Cartan matrix, see~\cite{georgi}.

\subsection{Wilson lines}

The presence of Wilson lines changes the modular invariance
condition~\cite{dhvw}. In addition to Eq.~(\ref{modinv}),
we have the condition for each Wilson line shift $a_i$,
$$
 (N \phi)^2 = (N V + N m_i a_i)^2 \quad (\mbox{mod } 2N).
$$
Here, by the point group action as well as
the lattice translation by
$m_i$, the Wilson-line-shifted vector is made
coincident to another shift vector.
Since we have no preference on the choice of reference
lattice axis, we can always set $m_i$ to be 1.
Therefore, one can show that this can be simply restated as,
\begin{equation}
\begin{split}
 N V \cdot a_i &= \mbox{integer}, \\
 N a_i \cdot a_j &= \mbox{integer}, \quad (i \ne j).
\end{split}
\end{equation}
This can be written in the Dynkin basis as,
$$
w_i \cdot A^{-1} s = 0 \quad \mbox{(mod } N),
$$
and
$$
w_i \cdot A^{-1} w_j = 0 \quad \mbox{(mod } N, i \ne j).
$$

In the presence of more than one Wilson line, these conditions
constrain the theory severely and reduce drastically
the number of possibilities.

In the $E_8 \times E_8^\prime$ heterotic string,
we can independently consider two sectors independently.
When we focus on the ``visible'' sector only, the
modular invariance condition is loosened, because we can put
the unwanted shift vector components into the hidden sector.
For the modular invariance, the full rank--16 group must be
considered for the condition (\ref{modinv}).
However, physics of the hidden sector is still important.
When we consider twisted sectors, generally multiplets are
hung on both sectors.

It is easily checked that these modular invariance conditions are
not spoiled by the automorphisms of the group. As seen in TABLE
\ref{e6furth}, therefore the same groups that come from the equivalent
breaking survive together if one survives.

\section{Classification scheme}

From the above construction, we can classify which
group emerges when we orbifold. The general strategy for
obtaining an orbifolded unbroken group is the following:
\begin{enumerate}
\item Determine the number of compact dimensions and the
orbifold which determines the order
$N$ of the shift.
\item Find all possible shift vector $v$  having the coefficient $s_i$
which satisfies Eq.(\ref{Kacond}) for given order $N$, for single
gauge group $E_8$.
\item The allowed shift vector $V$ for $E_8 \times E_8$ can be selected from
the combination of two shift vectors for single $E_8$ by requiring the
modular invariance condition (\ref{modinv}).
\item The surviving gauge group for the chosen shift vector can be read off
from the extended Dynkin diagram using the method explained in this paper.
\item Whenever we add Wilson line $a_i$, repeat the process from step 2,
  using Eq.(\ref{Kacond_sub}) this time.
\end{enumerate}

We note the merits of this procedure,
\begin{itemize}
\item
  Simple and intuitive.
\item
  Complete classificiation: We can take into
  account all the symmetries from a given group.
\item
  Easy to find the origin of a symmetry:
  We can see pictorially which symmetry comes
  from which group.
  From the example we presented earlier, the origin of
  $SU(3)^3$ from $E_6$ is easily seen.
\end{itemize}

Since the above procedure is easily implementable in the computer
program, we will present the classification tables of
unbroken groups with shift vectors and Wilson lines,
including the matter spectrum, in a separate
publication\cite{chk}.

In this paper, we set out the rules for finding out all the
unbroken gauge groups with all the possible shift vectors and
Wilson lines. The main point of the procedure we proposed is that
we can exhaust the Weyl reflection symmetry by requiring the shift
vectors and the Wilson lines to be the standard form for each
subgroup, which is still simple enough to read off the surviving
gauge group from the extended Dynkin diagram at the same time.
Since we obtain all the allowed gauge groups with shift vectors
and Wilson lines, the search for models with different sets of
chiral matters within a given gauge group is much more simplified
and tractable. It is even straight forward to identify the
untwisted and twisted matter representations for every possible
orbifold models, since we have done the classification of the
Weyl-non-equivalent shift vectors and Wilson lines, if needed by
the aid of the computer program\cite{chk}.


\begin{acknowledgments}
We thank
the Physikalisches Institut of the
Universit\"at Bonn for the hospitality extended to us
during our visit when this work was completed. This work
is supported in part by the KOSEF Sundo Grant(2002),
the BK21 program of Ministry of Education, and Korea
Research Foundation Grant No. KRF-PBRG-2002-070-C00022.
\end{acknowledgments}

\end{document}